# Time and the Unconscious Mind: A Brief Commentary

Julia Mossbridge, M.A., Ph.D.
Visiting Scholar, Department of Psychology, Northwestern University, Evanston, IL 60208
Founder and Research Director, Mossbridge Institute, LLC, Evanston, IL 60202
Please send correspondence to Julia Mossbridge: jmossbridge@gmail.com

Most of us think we know some basic facts about how time works. The facts we believe we know are based on a few intuitions about time, which are, in turn, based on our conscious waking experiences. As far as I can tell, these intuitions about time are something like this:
1) There is a physical world in which events occur,
2) These events are mirrored by our perceptual re-creation of them in essentially the same order in which they occur in the physical world,
3) This re-creation of events occurs in a linear order based on our conscious memory of them (e.g., event A is said to occur before event B if at some point we do remember event A but we don't yet remember event B, and at another point we remember both events),
4) Assuming we have good memories, what we remember has occurred in the past and what we don't remember but we can imagine might: a) never occur, b) occur when we are not conscious, or c) occur in the future.

These intuitions are excellent ones for understanding our conscious conception of ordered events. However, they do not tell us anything about how the non-conscious processes in our brains navigate events in time. Currently, neuroscientists assume that neural processes of which we are unaware, that is, non-conscious processes, create conscious awareness as a reflection of physical reality (Singer, 2015). Thus, if we wish to understand how events unfold in time in the physical world, we would do well to attempt to get some hints about how these events are navigated by non-conscious processes.

To understand non-conscious processing, we'd better first define consciousness. Here I define consciousness to be equivalent to its common neuroscientific definition, recently offered by Chris Koch, Chief Scientific Officer at the Allen Institute for Brain Science. "By consciousness, I mean the ability to feel something, anything -- whether it's the sensation of an azure-blue sky, a tooth ache, being sad, or worrying about the deadline two weeks from now" (Koch, 2012). Thus, non-conscious processing would be any neural processing of which we are not aware. It is critical to point out that because we are not aware of non-conscious processing and therefore have little intuitive understanding of it, our intuitions about time that are necessarily based on our conscious waking experiences will not necessarily apply to this non-conscious domain.

What we know about how non-conscious processes navigate physical events in time comes to us as shadows on a cave wall. That is, we can only informally and formally observe what we presume are the side effects of the non-conscious processing of events, and we can make inferences based on these observations. An example of an informal observation of what we presume to be such a side effect is the experience of déjà vu, in which we re-experience an event that we feel we have experienced before. Neuroscientists and experimental psychologists generally infer that déjà vu experiences arise from our non-conscious processes either mis-representing the original event so we believe incorrectly that the "past" event is the same as the current event, or misfiring to inappropriately given us a clear sense of familiarity even as we consciously experience an event for the first time (Brown, 2003). Either way, the assumption is that these non-conscious processes are mistaken in some way, because the standard arbiter of how events unfold in time seems to be our conscious waking experience of events. Regardless of the mechanisms underlying déjà vu, however, it is critical to step back from this assumption, as it is not at all clear that our conscious awareness accurately represents how events exist in the physical world. Of course, formal experiments on the influence of non-conscious processing of events are more critical than informal observations for our scientific understanding of how non-conscious processes navigate time. But before discussing formal experiments, it is important to first drive home the point that the version of events taking place over time in our waking conscious awareness is not necessarily, and not even probably, an accurate portrayal of events taking place in the physical world.

Simple logic provides the best evidence to support the realization that what seems to be occurring in our conscious awareness is probably not what is actually occurring in the physical world. Essentially, we do not need to know what is happening in the physical world; we only need to perceive what is relevant to our survival, and we need to perceive this in a way that guides our actions appropriately, but not necessarily in a way that accurately depicts every physical event. Perceptual psychologist Donald Hoffman has formalized this logic in several computational experiments in which he attempts to determine which of several types of automata "evolve" over time to dominate a simple environment (e.g., Hoffman, 2009; Hoffman & Singh 2012; Mark, Marion, & Hoffman, 2010). These automata have simple rules that both govern the information provided to their primitive nervous systems and that inform their actions. Regardless of the rules, his results reveal one consistent principle: automata that represent what is actually happening in the environment do not survive. They are beaten out in the evolutionary game by automata that do not accurately "perceive" the environment but instead perceive only several useful features of it, which are represented in simple, non-veridical ways. It turns out the energy required for accurate representation of the environment dooms a species to extinction. Meanwhile, who thrives? Species that get useful information about an environment without accurately representing what is really there.

The physical sciences support the idea that our intuitions based on conscious waking experiences of events do not give us accurate information about how events unfold in time. For example, Einstein' special theory of relativity illustrates that there can be no privileged "now" – an idea that, among other consequences, makes "past" and "future" observer-dependent so there is no absolute order of events when it comes to the physical world (Einstein, 1920). As a result, all events must co-exist in the physical world, and our conscious experience of traveling through time and discovering events one by one appears to be simply the most common way to experience them.

Experimental evidence in the psychology of perception also supports the contention that there is a mismatch between events in physical reality and what we experience in our conscious awareness (for review, see Hoffman, 2000). The illusion of motion from still images being flashed quickly in front of our eyes is one obvious example. Less obvious is the illusion of solidity of a wall, which we now know is the result of our perceptual systems being on a certain scale relative to the atoms in the wall. These are examples in which our conscious awareness has got it wrong in terms of its intuitions about the nature of the physical world.

As alluded to previously, there are multiple reasons to suspect that our conscious awareness is equally imperfect when it attempts to inform us about how time works. In fact, the accepted model of how events appear to unfold in time is woefully outdated. According to this model, *first* an event occurs in the physical world – in this case, we will assume the event is an appearance of a duck. *Second*, a partial representation of the duck is created using our non-conscious visual processing systems. *Third*, a full representation of the duck is constructed from partial representations and presented to our conscious awareness, which is considered to be the point of the whole process (e.g., Singer, 2015). In this model, creating an accurate representation of physical events occurs in an absolute, linear physical time. Take special note of these two features of the model in light of what we have discussed so far. We already know that an accurate representation of a physical event is not likely to be presented to conscious awareness, and we have known for decades that the absolute order of events is an illusion. Yet, the model persists.

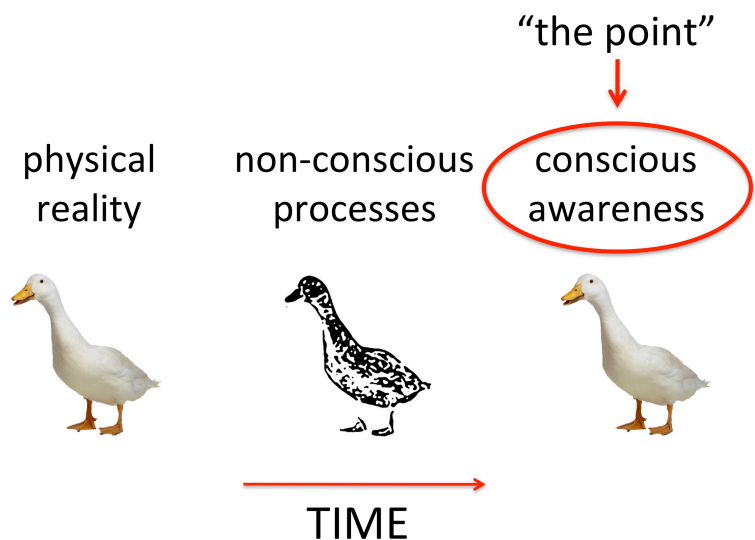

**Figure 1.** The current psychological/neuroscientific model of how events in time are processed by non-conscious awareness and reach conscious awareness. See text for details.

With what shall we replace this model? Before we attempt to build a new model, we ought to look at some additional experimental evidence that any model must take into account. Specifically, we will look at a well-controlled study examining non-conscious processing of information about the so-called "past" as well as a set of studies examining non-conscious processing of information about the so-called "future".

In an ingenious series of experiments designed to formally investigate non-conscious processing of events, Li et al. (2014) showed that non-conscious processes seem to integrate information about events that occur "farther back" in time than conscious awareness does. They asked participants to choose which of four smartphones they would purchase after giving the participants a list of features of each phone. Only two of these phones were of interest, because these were the phones for which there was a clear difference in the number of positive and negative features associated with them; I'll call these two, phone 1 and phone 2. The list of features was broken up into two sets, one set shown to participants before a break, and one set shown to participants after the break. The first set of features consisted of almost entirely positive features describing phone 1 and almost entirely negative features describing phone 2. After the break, the second set of features shown to participants was equivocal; neither phone was associated with more positive features than the other. Once all the features were shown to the participants, one half of the participants were asked to deliberate consciously about which phone was the best. About 50% of these deliberating participants chose phone 1 and the other 50% chose phone 2. Participants in the other group were asked to perform a completely unrelated task that demanded much of their attention – essentially allowing their non-conscious processing to work without any conscious demands on the question about the smartphone purchase. These participants overwhelmingly chose phone 1, which was the correct phone to choose if one integrated the entire list of features instead of just considering the features presented after the break (Li et al., 2014). These results and the control experiments described in that study suggest that in any new model, we build must take into account the ability of non-conscious processing to access information about events that conscious awareness does not usually access.

Now turning to examinations of the non-conscious processing of future events, we consider presentiment as an example. Multiple researchers have reported results from experiments testing the idea that physiological measures seem to be able to predict the emotional valence and/or intensity of events 2-15 seconds in the so-called "future" (e.g., Radin, 1997; Bierman & Scholte, 2002; Spottiswoode & May, 2003; McCraty et al., 2004a, b; Radin, 2004; May et al., 2005; Radin & Lobach, 2007; Radin & Borges, 2009). A meta-analysis examining many of these experiments published between 1978-2010 reports a significant overall, but small, effect (Mossbridge et al. 2012). These results suggest that at least some physiological processes, which are often not conscious, have access to events that seem to our conscious awareness to occur in the "future" (Mossbridge et al. 2012; Mossbridge et al. 2014). Pilot experiments using a pre-registered protocol and a smartphone application that gathers heartbeat data from hundreds of participants support this conclusion (see registry document for more information: http://www.koestler-parapsychology.psy.ed.ac.uk/Documents/KPU_Registry_1005.pdf). If we think clearly about what conclusion we can draw from these presentiment studies, we realize that it is the same conclusion as that drawn from the representative study examining non-conscious access to information presented in the so-called "past." Specifically, we can conclude that in any new model, we must take into account the data suggesting that non-conscious processing can access information about events that conscious awareness does not usually access.

We do not have to place events accessed by non-conscious processing into a "past" or "future." In fact, we cannot accurately do so, because we only know what "past" and "future" mean when we are talking about the intuitions arising from our conscious awareness of events. We have no idea where to place events in time outside of conscious awareness, so we should not attempt to do so. Working only with what we know, we can make the following claims:
1) Events occur in the physical world, and some are outside the awareness of consciousness.
2) Our non-conscious processing seems to access information about more events than conscious awareness.
3) Our non-conscious processing thus appears to have better access to physical reality than conscious awareness.
4) Our conscious awareness is provided information about events in a way that creates the appearance of a "past" that we remember somewhat hazily and a "future" that we do not remember at all.

These statements inform a candidate model for a new understanding of how events that seem to occur linearly in our waking conscious awareness arise from non-conscious processing.

In this model (Figure 2), we assume that it is unlikely that a real physical event is presented to conscious awareness as a fully accurate representation of physical reality. This assumption is schematized in Figure 2 by showing a strange creature in physical reality, which is always represented as a duck in our conscious awareness, by virtue of being presented to conscious awareness by non-conscious processes in a consistent "duck-like" way. We also assume that linear time and a privileged "now" that separates "past" from "future" is a construction that occurs in conscious awareness only. Non-conscious processes present a useful version of physical reality to conscious awareness, and for these processes a differentiation between "past" and "future" must be used to inform the presentation of reality to conscious awareness, but is not necessarily called upon when non-conscious processes are operating on their own. In sum, conscious awareness in this model gets demoted from *the point* of the whole process to a simple and incomplete *story* that, while it fails to reflect what is actually occurring in physical reality, still allows us to function in that reality.

# Events in time: a more accurate model?

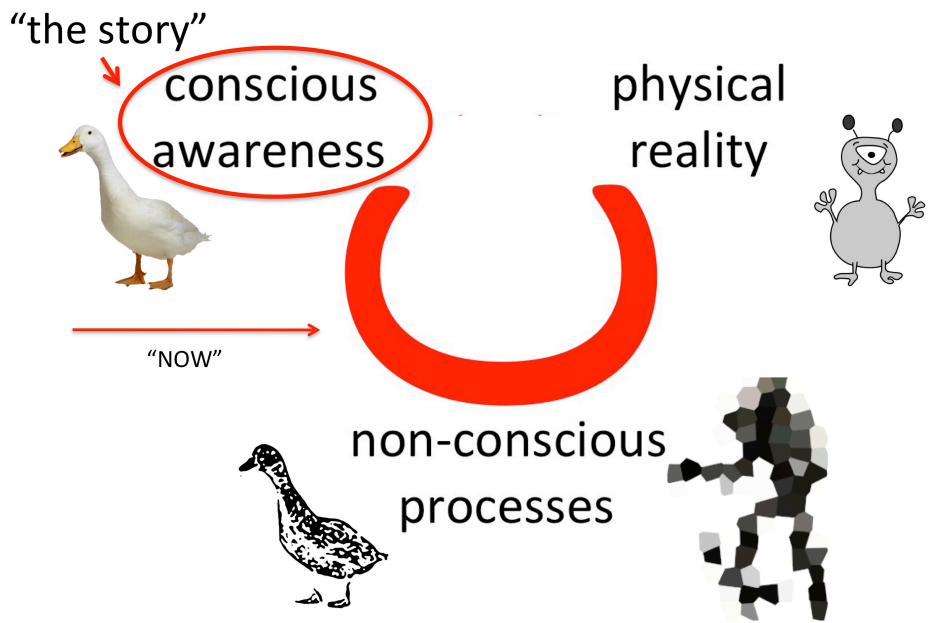

**Figure 2.** One candidate for what may be a more accurate model of how events in time are processed by non-conscious awareness and reach conscious awareness. See text for details.

If psychologists, neuroscientists, and philosophers could discard outdated models based on provably incorrect intuitions about how events unfold in physical reality (Figure 1) and begin to work with models that at least take into account what we think we know about physical reality and the illusion of time (Figure 2), we would be well on our way to integrating results from so-called mainstream and so-called parapsychological investigations. If we acknowledge the fallacies inherent in our experience-based intuitions about the physical world, including those about how time works, then the controversies that surround presentiment, precognition, telepathy, and psychokinesis might begin to fall away. Of course, if this happens we will have to put a lot more effort into how to read the shadows on the cave wall from which we infer that these non-conscious processes seem to know more than "we" do. This would be a happy outcome for many of us who are attempting to understand both consciousness and the non-conscious processes that are in such a close relationship with it.

**Acknowledgements**

This article was adapted from a webinar given in February 2015 to the Dutch Society for Psychical Research; many thanks to Richard Bierman, Eva Lobach, Dean Radin, and the audience of that webinar for comments that helped create this manuscript. The Bial Foundation provided a generous grant to support the work discussed in this commentary.

## References

Bierman D & Scholte H. 2002. A fMRI Brain Imaging study of presentiment. *Journal of ISLIS,* 20:380-389.

Brown, A. S. 2003. A review of the deja vu experience. *Psychological Bulletin*, 129:394–413.

Einstein, A. 1920. *Relativity: The special and general theory.* London, UK: Methuen & Co, Ltd. Translated by Robert W. Lawson.

Hoffman, D & Singh, M. 2012. Computational evolutionary perception. *Perception,* 41:1073-1091 (special issue in honor of David Marr).

Hoffman, D. 2000. *Visual intelligence: How we create what we see*. New York: W.W. Norton & Co.

Hoffman, D. 2009. The interface theory of perception: Natural selection drives true perception to swift extinction. In *Object categorization: Computer and human vision perspectives*, S. Dickinson, M. Tarr, A. Leonardis, B. Schiele (Eds.) Cambridge, UK: Cambridge University Press, 148–165.

Koch, C. 2012. Consciousness is everywhere. http://www.huffingtonpost.com/christof-koch/consciousness-is-everywhere_b_1784047.html (last accessed March 2, 2015).

Li J, Gao Q, Zhou J, Li X, Zhang M, Shen M. 2014. Bias or equality? Unconscious thought equally integrates temporally scattered information. *Consciousness & Cognition*, 25:77-87.

Mark, J, Marion, B & Hoffman, D. 2010. Natural selection and veridical perception. *Journal of Theoretical Biology*, 266:504–515.

May, EC, Paulinyi, T, Vassy, Z. 2005. Anomalous Anticipatory Skin Conductance Response to Acoustic Stimuli: Experimental Results and Speculation About a Mechanism. *Journal of Alternative & Complementary Medicine*, 11:695-702.

McCraty, R, Atkinson, M, Bradley, R.T. 2004a. Electrophysiological Evidence of Intuition: Part 2. A System-Wide Process? *Journal of Alternative & Complementary Medicine*, 10:133:143.

McCraty, R, Atkinson, M, Bradley, R.T. 2004b. Electrophysiological Evidence of Intuition: Part 1. The Surprising Role of the Heart. *Journal of Alternative & Complementary Medicine*, 10:325:336.

Mossbridge, J.A., Tressoldi, P., and Utts, J. 2012. Predictive anticipatory activity preceding seemingly unpredictable stimuli: A meta-analysis. *Frontiers in Psychology*, 3:390.

Mossbridge, J.A., Tressoldi, P., Utts, J., Ives, J.A., Radin, D. and Jonas, W.B. 2014. Predicting the unpredictable: critical analysis and practical implications of predictive anticipatory activity. *Frontiers in Human Neuroscience,* 8:146.

Radin, D. 2004. Electrodermal presentiments of future emotions. *Journal of Scientific Exploration*, 18:253-273.

Radin, D. & Borges, A. 2009. Intuition through time: what does the seer see? *Explore,* 5:200-211.

Radin. D. & Lobach, E. 2007. Toward understanding the placebo effect: Investigating a possible retrocausal factor. *Journal of Alternative and Complementary Medicine,* 13:733-739.

Singer, W. 2015. The Ongoing Search for the Neuronal Correlate of Consciousness. In T. Metzinger & J. M. Windt (Eds). *Open MIND:* 36(T). Frankfurt am Main: MIND Group.

Spottiswoode, S. & May, E. 2003. Skin conductance prestimulus response: Analyses, artifacts and a pilot study. *Journal of Scientific Exploration,* 17:617-641.